\documentclass[a4paper,12pt]{article} 
\usepackage{graphics}
\begin{document}
\setlength{\unitlength}{0.5mm}
\title{
Bounds on Integrals of the Wigner Function}
\author{ 
A.J. Bracken,$^{*}$
\& H.-D. Doebner\\
Arnold Sommerfeld Institute for Mathematical Physics, \\
TU Clausthal, \\
Leibnizstr. 10, \\
38678 Clausthal-Zellerfeld,\\
Germany\\  \\
and \\ \\J.G. Wood\\
Centre for Mathematical Physics,\\
Department of Mathematics, \\University of
Queensland, \\Brisbane 4072, Australia} 
\maketitle
\begin{abstract}
The integral 
of the Wigner function 
over a subregion of
the phase-space 
of a quantum system
may be less than zero or greater than one. 
It is shown that for 
systems
with one degree of freedom,
the problem of determining the best possible 
upper and lower bounds on 
such an integral, over all possible states,
reduces to the problem of finding the greatest and least 
eigenvalues of 
an hermitian operator corresponding to the
subregion.
The problem is
solved exactly in the case of 
an arbitrary elliptical region.   
These bounds provide 
checks on  
experimentally measured quasiprobability distributions.
\end{abstract}


The Wigner 
function 
has been much studied since
its introduction
\cite{Wigner},   
not only in the context of 
quantum physics 
\cite
{Baker}, 
but also in
signal processing \cite{signals}.  
For a quantum system in a pure state, the Wigner function 
carries the same information as the wavefunction, up to an
unimportant constant phase.  In the case of a mixed state, 
it carries the same information as the density operator.

An important property of the Wigner function, one of several 
properties which distinguish it from  classical probability densities, 
is that
its integral
over  a given subregion of phase-space
may be negative or greater than one.   
Quasiprobability distributions
which, according to quantum theory, 
correspond to Wigner functions, 
have been measured 
in recent experiments, 
for a variety of states of light and matter 
\cite{light,molecules,Leibfried,matter,Leonhardt,general,reviews},
and negative values have indeed been observed.
These  
experiments
are probing  the basic structure and predictions of 
quantum mechanics in a new way,   
and the prospect of increasingly accurate experiments of this
type adds greatly to the interest in, and importance of, the
theory of the Wigner function.

We 
consider the problem of determining the best possible
bounds on the integral of the  Wigner function over a given
subregion of the phase-plane of any system with one degree of
freedom.  
We show for any subregion of a rather general type that 
this problem reduces to the problem of finding the greatest and
least eigenvalues of an hermitian Fredholm integral operator
corresponding to that subregion.    
The problem is found to be exactly solvable
for any elliptical or annular
subregion, and the bounds are given explicitly in the 
case of the ellipse.  
These best possible
bounds provide new information about the structure of the  
Wigner function,  differing  
from known results such
as  best possible 
bounds 
(\ref{Wbounds})
on the values of the Wigner function
itself, 
bounds on integrals of powers of the function
\cite{Price}, or bounds on various
moments of the function \cite{deBruijn}.  In particular, the
new bounds
determine the degree to which the integral of any Wigner
function over an elliptical  
subregion of the phase-plane can lie outside
the interval $[0,1]$ which applies to classical densities. 
In principle, they therefore provide
checks on 
experiments of the type
to which we have referred, because
they must be respected by any
measured quasiprobability distribution consistent with 
quantum
mechanics.       

As we shall show, appropriately chosen  
oscillator stationary states (or single frequency light modes)
lead  
{\em theoretically}
to the exact attainment of  
these upper and lower bounds.
Such states 
are perhaps the easiest to establish
experimentally,
and it is just such  states for
which quasiprobability distributions 
have been measured in some of the
experiments mentioned above \cite{Leibfried}.   

In what follows, we consider  
systems with one degree of freedom,
with a Cartesian coordinate $q$ and its
conjugate momentum $p$\,.
Our results refer to the Wigner function 
considered at a particular
instant, and 
are therefore independent of 
any particular  dynamics.
We work in dimensionless variables. Appropriate
dimensional factors will appear in what follows
if each coordinate $q$ there is replaced by
$q/L$, each momentum $p$ by $Lp/\hbar$,  
each wavefunction $\psi$ by $L^{1/2}\psi$, each phase space
area $A$ by $A/\hbar$, and 
each Wigner function $W$  by $\hbar W$, where
$L$
is a suitable constant with dimensions of a length.    

Given a normalized wavefunction $\psi$ corresponding to a  
pure state $\vert\psi\rangle$,
the Wigner function is defined as 
\begin{equation}
W_{\psi}(q,p)=\frac{1}{\pi}\int_{-\infty}^{\infty}
\psi^{*}(q+x)
\psi(q-x)
e^{i2px}\,dx\,.
\label{Wdef}
\end{equation}
Then \cite{Wigner}
\begin{equation} 
\int_{\Gamma}W_{\psi}\,dq\,dp =1\,,\quad
\int_{\Gamma}
[W_{\psi}]^2\,dq\,dp =
\frac{1}{2\pi}
\,,  
\label{W_ints}
\end{equation} 
where $\Gamma$ denotes the $(q,p)$ phase-plane.   

For a mixed state, the density operator $\rho$
is  positive-definite and 
hermitian with unit trace, and typically
can be resolved in
the form 
\begin{equation}
\rho = \sum_i p_i \vert \psi _i\rangle \langle \psi_i\vert 
\,,\quad p_i> 0\,,
\quad \sum _i p_i =1\, ,
\label{mixedRho}
\end{equation}
where the states $\vert\psi_i\rangle$ are orthonormal.   
The corresponding Wigner function has the form 
\begin{equation}
W_{\rho}= \sum_i p_i W_{\psi _i}\,,
\label{mixedW}
\end{equation}
where $W_{\psi _i}$ is the Wigner function corresponding to the
pure state $\vert\psi_i\rangle$.  
More generally, the sum in (\ref{mixedRho}) and (\ref{mixedW})
could be  replaced in part or whole by an integral, 
but this does not significantly affect the 
argument of the next paragraph. 

It follows from (\ref{mixedRho}) and  
(\ref{mixedW}) that any bound on 
the Wigner function, 
or on its integral 
over a given subregion $S$ of $\Gamma$, 
must hold for all possible mixed states
if it holds for
all possible 
pure states. 
For example, if 
\begin{equation}
\int_S W_{\psi} (q,p)\,dq\,dp > L \quad{\rm for \,\,\, all} \quad\psi\,,
\end{equation}
then for any $\rho$ as in (\ref{mixedRho}), 
\begin{eqnarray}
\int_S W_{\rho}(q,p)\,dq\,dp &=&\sum_i p_i\int_S W_{\psi _i}
(q,p)\,dq\,dp \nonumber\\
&>& \sum _i p_i L = L\,.
\end{eqnarray}
Since a pure state can be regarded as a
limiting case of a mixed state, it then follows that
{\em best possible} upper
and lower bounds on  
the Wigner function or its integral, 
when considered over all pure states,
must also be best possible upper or lower bounds
when considered over all
mixed states, although a bound that is {\em attainable} over
pure states may not in general be attainable over mixed states.
Bearing this in mind, we restrict attention in what follows to
pure states.  

Best possible bounds on the Wigner function itself are known
\cite{Baker}:
\begin{equation}
-\frac{1}{\pi}\,
\leq
W_{\psi} (q,p)  \leq\, 
\frac{1}{\pi}
\label{Wbounds}
\end{equation}
for all normalized 
$\psi$,  
for all 
$(q,p)\in\Gamma$.
It is easily seen that $W_{\psi} = \pm 1/\pi$ at
the point $(q,p)$ if and only if 
\begin{equation}
\psi (q-x)e^{ipx} = \pm \psi(q+x)e^{-ipx}\quad{\rm for\,\,\, all
}\,\,x\,.
\label{Wattain}
\end{equation}

The problem of interest here is to find best possible bounds on
the `quasiprobability functional' corresponding to the subregion
$S$, defined as 
\begin{eqnarray}  
Q_S [W_{\psi}]&=& \int_S W_{\psi}(q,p)\,dq\,dp 
\nonumber\\
&=&\int _{\Gamma} \chi _S (q,p) W_{\psi} (q,p)\,dq\,dp
\,,\label{mass}
\end{eqnarray}  
where 
$\chi _S$ is 
the function with the value
$1$ on $S$, and the value 
$0$ on the complement of $S$.  

It follows at once from (\ref{Wbounds}) and (\ref{mass})
that
\begin{equation} 
-
\frac{A_S}{\pi}
\,\leq\,
Q_S[W_{\psi}]
\,\leq\,
\frac{A_S}{\pi}\,,
\label{Ibounds1}
\end{equation}
where 
$A_S = \int_S\,dq\,dp$ 
is the area of $S$.  

In order to obtain stronger bounds than (\ref{Ibounds1}), recall  
that each
real-valued function $T(q,p)$ on $\Gamma$
can be associated with an 
hermitian operator ${\hat T}$ 
such that 
\begin{equation}
(\psi, {\hat T}\psi)=\int_{\Gamma} T(q,p) W_{\psi}(q,p) \,dq\,dp\,,
\label{corresp}
\end{equation} 
where $(\psi_1,\psi_2)$ is usual scalar product of
wavefunctions.
Here ${\hat T}$ can always be written as a
Fredholm integral operator,  
\begin{equation} 
({\hat T}\psi)(x) =
\int_{-\infty}^{\infty}
K_T (x,y)\psi(y)\,dy\,,
\label{Aoperator}
\end{equation} 
with hermitian kernel given in terms 
of the real-valued function  $T(q,p)$ as 
\begin{equation} 
K_T(x,y)=
\frac{1}{2\pi}
\int_{-\infty}^{\infty} 
T((x+y)/2,p)\,e^{ip(x-y)}\,dp\,.
\label{Akernel}
\end{equation} 

Consider now the case when $T(q,p)=\chi_S(q,p)$.   Comparison of
(\ref{mass}) and (\ref{corresp}) shows that 
\begin{eqnarray} 
Q_S[W_{\psi}]&=&(\psi,{\hat K}_S\psi)\,,
\label{expec}\\
\nonumber\\   
({\hat K}_S\psi)(x)&=&
\int_{-\infty}^{\infty} 
K_S(x,y)\psi(y)\,dy \,,
\label{Kop}\\
\nonumber\\
K_S(x,y)&=&
\frac{1}{2\pi}
\int_{-\infty}^{\infty} 
\chi_S ((x+y)/2,p)e^{ip(x-y)}\,dp \,.
\label{Kkernel1}
\end{eqnarray} 
It follows at once from (\ref{expec}) 
that the extremal values of $Q_S[W_{\psi}]$ 
are determined by the eigenvalue
problem 
${\hat K}_S \psi = \lambda \psi$
with ${\hat K}_S$ as in (\ref{Kop}). In particular, 
\begin{equation} 
\inf\, Q_S = 
\lambda_{min}\,,\quad
\sup Q_S = 
\lambda_{max}\,,
\label{supinf}
\end{equation} 
where 
$\lambda_{min}$ and $\lambda_{max}$ are the least and greatest  
eigenvalues of ${\hat K}_S$ respectively (or more generally, the
infimum and supremum of the spectrum of ${\hat K}_S$). 
Thus the 
problem of interest now becomes the determination of 
$\lambda_{min}$ and $\lambda_{max}$.    

In order to proceed, 
suppose that the subregion 
$S\subset\Gamma$ has the general form shown in Fig. 1.
\begin{figure}[h]
\begin{center} 
\begin{picture}(100,120)
\thinlines

\put(0,50){\vector(1,0){120}}
\put(50,0){\vector(0,1){100}}

\put(54,96){$p$}
\put(113,44){$q$}

\put(50,50){\circle{6}}

\linethickness{1pt}
\bezier{50}(40,70)(50,90)(60,90)
\bezier{50}(60,90)(62,90)(65,85)
\bezier{50}(65,85)(68,80)(70,80)
\bezier{50}(70,80)(72,80)(75,83)  
\bezier{50}(75,83)(78,86)(82,86)
\bezier{50}(82,86)(86,86)(98,78)
\bezier{50}(40,70)(44,56)(46,56)
\bezier{50}(46,56)(52,56)(55,62)
\bezier{50}(55,62)(57,68)(62,68)
\bezier{50}(62,68)(66,68)(68,54)
\bezier{50}(68,54)(70,40)(74,40)
\bezier{50}(74,40)(76,40)(78,56)
\bezier{50}(78,56)(80,72)(83,72)
\bezier{50}(83,72)(86,72)(88,67)
\bezier{50}(88,67)(90,62)(93,62)
\bezier{50}(93,62)(95,62)(98,78)

\put(56,74){$S$}

\multiput(40,51)(0,4){5}{\line(0,1){2}}
\multiput(98,51)(0,4){7}{\line(0,1){2}}

\put(38,44){$b$}
\put(96,44){$c$}

\put(36,80){\line(-3,2){16}}
\put(32,88){\line(1,-2){4}}
\put(32,88){\vector(3,-2){12}}

\put(6,94){$p=F_2 (q)$}

\put(34,30){\vector(3,1){36}}
\put(34,30){\line(5,-1){10}}
\put(29,23){\line(3,1){15}}

\put(10,14){$p=F_1(q)$}


\end{picture}
\end{center}
\end{figure}
\vskip5mm
\noindent
\centerline{Fig.1. A typical region $S$ in the phase-plane.}
\vskip5mm
\noindent
Here $F_1$ and $F_2$ are real-valued 
functions defined for $b\leq q\leq c$, and satisfying
$F_1 (b) = F_2 (b)$, 
$F_1(c)=F_2(c)$, and $F_2(q)\geq F_1(q)$ for $b<q<c$. 
Each function need only be piecewise continuous, and 
$b= -\infty$ and/or $c=\infty$ is allowed.
For such a subregion, the characteristic function has the form
\begin{equation}
\chi_S (q,p)
=\left\{\begin{array}{lr}
1& \quad b<q<c\,,\quad F_1(q)<p<F_2(q)\\
0& \quad {\rm otherwise}\,,\quad\qquad\qquad\qquad\qquad 
\end{array}
\right.  
\label{chiform}
\end{equation} 
and the kernel 
(\ref{Kkernel1}) becomes
\begin{eqnarray} 
K_S(x,y)&=&
\frac{1}{2\pi}
\int_{F_1(\frac{x+y}{2})}^{F_2(\frac{x+y}{2})} 
e^{ip(x-y)}\,dp 
\nonumber\\
\nonumber\\
&=&
\frac{e^{i(x-y)F_2(\frac{x+y}{2})}
-e^{i(x-y)F_1(\frac{x+y}{2})}}{2\pi i (x-y)}\,,
\label{Kerform}
\end{eqnarray} 
for $2b<(x+y)<2c$, and $0$ otherwise. 
Note that the singularity at $x=y$ is only apparent.
Then (\ref{Kop}) becomes
\begin{eqnarray} 
({\hat K}_S\psi)(x)= \quad\qquad\qquad\qquad\qquad\qquad
\nonumber \\
\nonumber \\
\,\,\int_{2b-x}^{2c-x} 
\frac{e^{i(x-y)F_2(\frac{x+y}{2})}
-e^{i(x-y)F_1(\frac{x+y}{2})}}{2\pi i (x-y)}\,
\psi(y)\,dy\,.
\label{Kop2}
\end{eqnarray} 

More generally, the subregion $S$ may consist of 
several nonintersecting parts $S_1$, $S_2$, $\dots$
of the same general type, even on overlapping 
$q$-intervals. It is easily seen that in such a case
${\hat K}_S = {\hat K}_{S_1} + {\hat K}_{S_2} + \dots$
However, in general 
$[{\hat K}_{S_1}\,,{\hat K}_{S_2}]\neq0,$
{\em etc.}, so that the bounds associated 
with different subregions cannot  be added. 

Note also that the extremal values of $Q_S$ and $Q_{S'}$ 
are the same if $S$ is transformed into $S'$  
by a canonical 
transformation of $\Gamma$ 
of the form  
\begin{equation} 
q' = \alpha q + \beta p + \gamma\,,\quad p'=\mu p +\nu q + \rho\,, 
\label{canon}
\end{equation} 
where $\alpha$, $\beta$, $\gamma$, 
$\mu$, $\nu$ and $\rho$ are real constants satisfying
$\alpha \mu - \beta \nu =1$.
In particular, the case of any circular or elliptical region
of area $\pi a^2$
can be reduced to the case of a circular disk of radius $a$,
centred at the origin.   

In this case, the operator ${\hat K}_S$ (let ${\hat K}_a$ denote it now)
is given from (\ref{Kop2}) by
\begin{eqnarray} 
({\hat K}_a\psi)(x)= \quad\qquad\qquad\qquad\qquad\qquad
\nonumber\\
\nonumber\\
\int_{-2a-x}^{2a-x} \frac{\sin[(x-y)\sqrt{a^2 - (x+y)^2/4}
\,]}{\pi (x-y)}\,\psi(y)\,dy\,, 
\label{circle1}
\end{eqnarray}
for $-\infty<x<\infty$,    
and it is not hard to check that 
${\hat K}_a$ 
commutes with the simple harmonic
oscillator Hamiltonian operator ${\hat H}$ defined by 
\begin{equation} 
{\hat H}\,\psi (x) = - \frac{d^2\psi (x)}{dx^2} + x^2\psi (x)\,.
\label{SHO}
\end{equation} 
This is explained by the fact that $\hat H$ generates
transformations of the wavefunction corresponding to rotations
in the phase-plane, which leave the disk invariant.
It follows that for {\em every} value of $a$
the eigenfunctions of ${\hat K}_a$ are
the
oscillator eigenfunctions
\begin{equation} 
\psi _n (x) = H_n (x)e^{-x^2/2}\,,\quad n=0,1,\dots
\label{osc_efns}
\end{equation} 
where $H_n$ is the Hermite polynomial \cite{Abram}.    

According to (\ref{expec}), 
the eigenvalue $\lambda _n(a)$ of ${\hat K}_a$ corresponding to 
the eigenfunction 
(\ref{osc_efns}),  
must equal the total 
quasiprobability on the disk of radius $a$, 
as determined by  the Wigner
function $W_n$ (say) corresponding to that eigenfunction.    
Since it is known \cite{Baker,deBruijn} that 
\begin{equation}  
W_n(q,p) = (-1)^n\pi ^{-1} L_n (2[p^2 + q^2])e^{-(p^2 + q^2)}\,,
\label{SHO_W}
\end{equation}
where $L_n$ is the Laguerre polynomial \cite{Abram}, 
it follows that
\begin{equation} 
\lambda _n (a) = (-1)^n \int _0^{a^2}L_n(2u)\,e^{-u}\,du\,.
\label{lambdaform}
\end{equation} 
Thus $\lambda_0 (a)= 1-e^{-a^2}$, $\lambda_1 (a) =
1-(1+2a^2)e^{-a^2}$, $\lambda _2 (a) = 1-(1+2a^4)e^{-a^2}$, 
$\lambda _3 (a)= 1-(1+2a^2 -2a^4+\frac{4}{3}a^6)e^{-a^2}$, {\em
etc.}   

\noindent
{\em In summary:} 
\begin{eqnarray}  
\int_{-2a-x}^{2a-x} \frac{\sin[(x-y)\sqrt{a^2 - 
(x+y)^2/4}\,]}{\pi (x-y)}\,\psi _n(y)\,dy 
\nonumber\\
\nonumber\\
\qquad\qquad\qquad\qquad= 
\lambda _n (a) \psi _n (x)\,,
\label{summary}
\end{eqnarray} 
with $\psi _n$ as in (\ref{osc_efns}) and $\lambda _n $ as in
(\ref{lambdaform}).

Fig. 2 shows the graphs of $\lambda_n$ {\em versus} $a$ for
$n=0,\, 1,\, 2,\, 3$, and also the graphs of $\lambda_{max}$ and
$\lambda_{min}$ (bold lines).    Note that
$\lambda_{max}(a)=\lambda _0 (a) = 1- e^{-a^2}$, whereas
the graph of $\lambda_{min}$ has the peculiar scalloped shape
shown, because 
$\lambda _{min} (a)=\lambda _1 (a)$ for $0\leq a<a_1$, 
$\lambda _{min} (a)=\lambda _2 (a)$ for $a_1\leq a<a_2$, 
{\em etc.}, where $a_1$ is the greatest value of $a$ 
at which $\lambda _1
(a)=\lambda _2 (a)$, $a_2$ is the greatest value of $a$ at 
which $\lambda _2 (a) =
\lambda _3 (a)$, {\em etc.}
Thus $a_1=1$, $a_2=\sqrt{(3+\sqrt{3})/2}$, {\em etc.}
\scalebox{0.7}{
\includegraphics{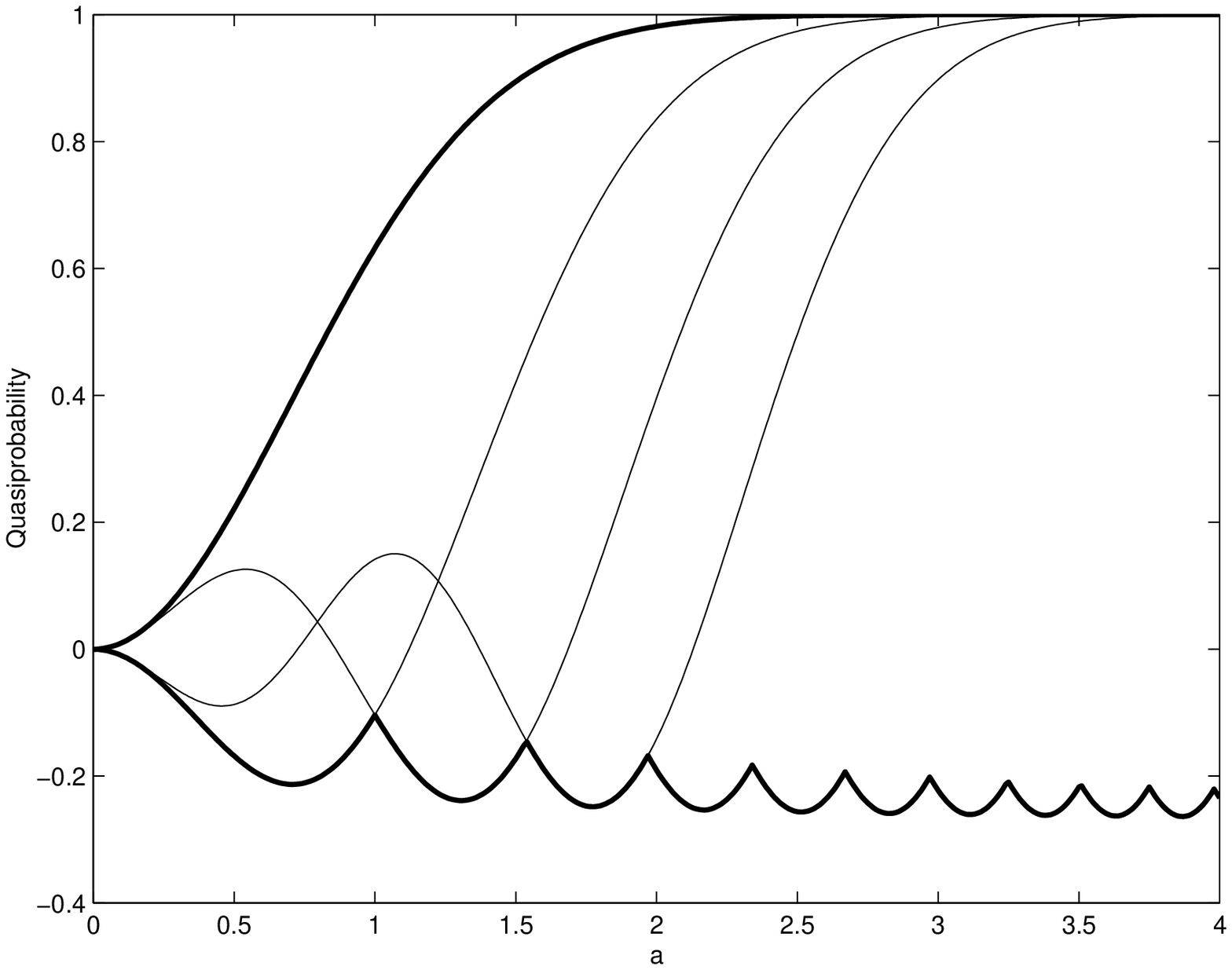}
}
\vskip5mm
\noindent
Fig. 2. Left to right: graphs of $\lambda _n$ for $n=0$, $1$,
$2$, $3$, and also of $\lambda_{max}$, $\lambda_{min}$ (bold lines). 
\vskip5mm
\noindent
With the introduction of  the appropriate dimensional factors,
the result is 
that the integral of any pure-state or mixed-state
Wigner function over 
any circular or elliptical region with area $\pi a^2\hbar$
in the phase-plane, 
lies in the interval 
$[\lambda_{min}(a),\lambda_{max}(a)]$, in contrast to the
integral of any 
classical density, which lies in $[0,1]$.  According to
quantum mechanics, any 
quasiprobability distribution
determined 
by quantum tomography
(in particular) 
is described by a Wigner function
\cite{Leonhardt,general,reviews}.  
For such a distribution, 
the quasiprobability on disks of various radii, centred on
regions where the distribution is most negative, for example,
could be
estimated and checked for consistency
against the theoretical bounds.
Of course, experimental data are inevitably subject to noise for
various reasons.   While there are known techniques to allow for
noise in the reconstruction of densities from more primitive
data 
\cite{Leonhardt,Simonoff}, this would  
obviously limit the
power of the proposed check.
A more subtle complication is that reconstruction
algorithms may 
invoke quantum mechanical arguments
\cite{Leonhardt,general,reviews}, and 
any check would 
be satisfied trivially in a given case
if these arguments {\em forced} a reconstructed
density to satisy the theoretical bounds.
Any given reconstruction algorithm
would have to be analysed carefully in this regard  
to ensure that a check was meaningful.   

If these difficulties could be overcome, might the proposed check 
be elevated to the level of a 
test of quantum mechanics itself?  The  assumptions
underlying the theory of the Wigner function are very few:  the
linear vector space of states, the Born interpretation, and
the conjugate relations between coordinates and momenta.
However, quantum mechanics has been so well-tested at the
energy scales of the present experiments that any violations, if
indeed there are any, 
must surely be exceedingly small, and very probably beyond
present capabilities of resolution amidst noise.

Eigenvalue problems corresponding to other shapes
such as squares
and triangles are easily formulated, but  
do not seem to be exactly solvable.  They could be
tackled numerically. 
Exact results for disks can be extended
to the case of an annular region
(and more generally the case of several concentric annuli),
because the operators ${\hat K}_a$ commute for
different $a$, and have common eigenfunctions.
This
may be particularly useful in checking distributions determined
by the `ring method' \cite{Leibfried}. 

These ideas can be extended
to systems with more degrees of
freedom, and to systems with spin.

\vskip5mm
Thanks are due to J.A. Belward, R. Chakrabarti, 
G.A. Chandler, D. Ellinas,  W.P.
Schleich and referees of a preliminary version 
for helpful
comments.

\vskip5mm
\noindent 

\end{document}